\documentclass[12pt]{iopart}

\usepackage{graphicx}
\usepackage{amsfonts}
\usepackage{xcolor}
\usepackage{cite}
\usepackage{subfigure}

\usepackage{bm}
\usepackage{epstopdf}

\begin{document}

\title[Gross-Pitaevskii validity  in reduced dimensions]{Validity of Gross-Pitaevskii solutions of harmonically confined BEC gases in reduced dimensions}

\author{R. Zamora-Zamora,$^{1} $ G.A. Dom\'{\i}nguez-Castro,$^{1}$ C. Trallero-Giner,$^{2,3}$ R. Paredes,$^{1}$ and V. Romero-Roch\'{\i}n$^{1}$}

\address{ $^{1}$ Instituto de F\'{\i}sica, Universidad
Nacional Aut\'onoma de M\'exico, Apartado Postal 20-364, M\'exico.
D.F. 01000, Mexico. }
\address{ $^{2}$ Department of Theoretical Physics, Havana University, Havana 10400, Cuba}
\address{ $^{3}$CLAF - Centro Latino-Americano de F\'{\i}sica, Avenida Venceslau Braz,
71, Fundos, 22290-140, Rio de Janeiro, RJ, Brasil}

\ead{romero@fisica.unam.mx}
\vspace{10pt}
\begin{indented}
\item[]
\end{indented}

\begin{abstract}
By exact numerical solutions of the Gross-Pitaevskii (GP) equation in 3D, we assess the validity of 1D and 2D approximations in the study of Bose-Einstein condensates confined in harmonic trap potentials. Typically, these approximations are performed when one or more of the harmonic frequencies are much greater than the remaining ones, using arguments based on the adiabatic evolution of the initial approximated state. Deviations from the 3D solution are evaluated as a function of both the effective interaction strength and the ratio between the trap frequencies that define the reduced dimension where the condensate is confined. The observables analyzed are both of stationary and dynamical character, namely, the chemical potential, the wave function profiles, and the time evolution of the approximated 1D and 2D stationary states, considered as initial states in the 3D GP equation. Our study, besides setting quantitative limits on approximations previously developed, should be useful in actual experimental studies where quasi-1D and quasi-2D conditions are assumed. From a qualitative perspective, 1D and 2D approximations certainly become valid when the anisotropy is large, but in addition the interaction strength needs to be above a certain threshold.
\end{abstract}

\section{Introduction}
Dimensionality plays a determinant role in the occurrence or not of different physical phenomena in many-body systems \cite{Mermin, Hohenberg, Kosterlitz, Castro, Wilder}. Bose-Einstein condensation (BEC) in ultracold gases of bosonic atoms, the subject of our attention here, is not an exception,  and most of the attention and success of the current ultracold matter experiments is directly related to the fact that these systems, composed of a macroscopic number of atoms and being in the degenerated regime, behave analogously to condensed matter systems. Even more, interparticle interactions and inhomogeneous potentials of variable dimensionality can be set in those ultracold quantum fluids by means of external fields to produce particular energy landscapes \cite{Choi, Bernhard, Potter, Vosk, Kjall, Pal, Hadzibabic, Chomaz}. Such experimental capabilities allow to explore the physics of a wide range of phenomena, for instance phase transitions in inhomogeneous potentials \cite{Arnold, Houbiers,Gerbier,Romero,Davis,Sandoval-Figueroa,Zobay}, transition to localization induced by either static or dynamic disorder \cite{Choi}, topological defects such as vortices \cite{Zamora1,Leanhardt, Abo-Shaeer}  and Skyrmions \cite{Choi2,Zamora2,Kawakami,Battye}, quantum turbulence, formation of magnetic domains in spinor condensates and relaxation to equilibrium versus eigenstate thermalization hypothesis, among others. Indubitably, as it is observed and predicted, dimensionality and interparticle interactions play a key role in the occurrence of the mentioned phenomena. Therefore, a central aspect to quantum simulate a given scenario is to establish the range of parameters which support both, experimental and theoretical approaches in 1D and 2D.\\

Bose-Einstein condensation in ultracold fluids has the peculiarity that, due to atomic interactions and their spatial confinement by inhomogeneous traps, they can become  superfluids in arbitrary dimensions. In those very low temperatures and diluted regimes, the mean field Gross-Pitaevskii equation (GP) provides an optimal scheme to describe those gases confined in inhomogeneous traps in all dimensions, and one can write down corresponding GP equations to describe them. A key and evident observation, however, is that all the condensates produced in the laboratory are in 3D space. Thus, the reduction of dimensionality is achieved by changing the geometry and the space where the atoms are confined, modifying and having a determinant effect on the effective resulting interactions. Typically, based on simple physical intuition, a reduction of dimensionality is achieved by constraining the dynamics of the condensate to fewer spatial degrees of freedom. This is enabled by reducing the spatial extent of the motion of the atoms in one or more directions. To be specific, assuming that the inhomogeneous trap is an anisotropic harmonic oscillator with 3D frequencies $\omega_x$, $\omega_y$ and $\omega_z$, then, if one or two of those frequencies become too large, the motion occurs in the remaining ones. Thus, one simply considers the dynamics in those coordinates, writes down the appropriate GP equation, and neglects the other ones. This procedure, however, requires a theoretical justification that both yields the correct equation in the reduced dimension and sets their bounds of validity. This theoretical procedure, of how to conduct such a dimension reduction, has been the subject of several studies, \cite{Lieb,Reatto,Mateo,Bao,Bao2,Trallero,Petrov, Pethick,Chiofalo,Nicolin} with the most important result that the atomic interaction strength becomes modified in 1D and 2D with respect to the original interaction in 3D, usually being scaled by the length of the oscillator in the constrained direction or directions.  In some of those cases \cite{Reatto,Mateo}, however, the reduced equations show a non-linearity different from the usual one in GP equations. This is certainly interesting and relevant and we shall also discuss these approximations in the light of the full 3D solution. The validity, as expected from those asymptotic procedures, should be obtained in the limit of arbitrarily large anisotropy. However, actual experiments are rarely in those asymptotic limits and, as the primary motivation, the purpose of the present study is the realization of a systematic and rigorous numerical analysis of the ensued 1D and 2D versions compared with the actual 3D descriptions of the GP equation describing a BE condensate at zero temperature.\\

We proceed by, first, briefly describing the two main different approaches \cite{Bao,Trallero} that deal with the reduction of dimensionality of the 3D GP equation to 2D and 1D effective equations. While differing in the precise step where the reduction is enabled, both procedures are based on a variational scheme followed by an adiabatic approximation which assumes that the dynamics is spatially decoupled; in this way, the component of the gas in the squeezed or small dimension or dimensions stays in its non-interacting ground state, and the atomic collisions, represented by non-linear terms, take place in the desired reduced dimension. Then, separately for the 1D and 2D cases, we perform four validity tests, three stationary and one dynamical, to compare the actual 3D solution with the approximated ones. These stationary and dynamical properties are obtained by an extensive set of numerical calculations as a function of the effective coupling interaction and of the ratio of the trap frequencies defining the reduction of dimensions in each case. Our calculations are parallel state-of-the-art computations with GPU processors, for which we provide the necessary technical details to allow for the reproducibility of our results and their practical use. As we shall conclude very generally, the true achievement of physics in reduced dimensions requires not only a large anisotropic confinement but large enough atomic interaction strengths, and we provide the required bounds. Some words of caution, however, are in order before we embark in those conclusions. The 3D GP equation is a non-linear equation and, as such, strictly speaking cannot be decoupled in transverse and longitudinal variables, namely, its solution  cannot be factorized into transverse and longitudinal components. The variational methods, nevertheless, based on defining a certain ansatz with variational parameters in the functional energy or Hamiltonian, allow to get the ``effective'' 1D and 2D ordinary differential equations, which describe the corresponding static and dynamic behavior. In this regard, it is known that variational methods do not necessarily guarantee that, say, while the chemical potential, being a stationary property, is accurately described, the corresponding condensate wavefunction is also correctly obtained; see Ref. \cite{Carretero} for a discussion of variational methods in GP equations. In the light of this observation, it is certainly of interest to study both properties, dynamical and stationary, and separately assert its validity, being careful of not dismissing partial agreement between the approximated 1D and 2D solutions and the full 3D version. 
And also, as we will discuss in the final section of this article, there emerges an interesting and relevant observation concerning the underlying explanation of the achievement of the dimension reduction, at least within the framework of the GP formalism, namely, that the interplay between the effective non-linear interactions and the {\it relative} anisotropic confinement
allows the dimension reduction rather than their absolute tightening.

\section{From 3D to 2D and 1D Gross-Pitaevskii descriptions}

The 3D Gross Pitaevskii equation that describes a Bose-Einstein condensate of $N$ atoms confined in a harmonic potential is
\begin{eqnarray}
 i\hbar \frac{\partial}{\partial t} \psi_{3D} (\mathrm{\bf r},t)&=&\left[-\frac{\hbar^2 }{2m} \nabla^2
 +\frac{1}{2}m (\omega_x^2 x^2+ \omega_y^2 y^2 + \omega_z^2 z^2) \right.  \nonumber  \\
  &&\quad\left. {} + \frac{4 \pi \hbar^2 N a}{m} |\psi_{3D} (\mathrm{\bf r})|^2  \right] \psi_{3D} (\mathrm{\bf r},t),   \label{GP3D}
\end{eqnarray}
where $a$ is the $s-$wave scattering length $a$ of the two-body interaction and $m$ the atom mass. The trap frequencies are labeled as $\omega_i$ along the $i = x,y$ and $z$ directions, respectively. Depending on the size of these frequencies the cloud of the condensate gas can be either an spheroid in 3D space, a disk lying in 2D, or a cigar shaped cloud in 1D. As we shall see in the next subsections, based on reasonable physical assumptions, if one or more frequencies along different axes are much larger with respect to the others, say, $\omega_x = \omega_y \gg \omega_z$, or $\omega_z \gg \omega_x= \omega_y$, the ``energy levels'' along different directions become so separated that, at arbitrarily low temperatures, the system decouples in the sense that the dynamics of the condensate is developed either, parallel to $z$-axis yielding a quasi 1D condensate, or in the $x-y$ plane for a 2D disk shaped condensate, while the other transverse degrees of freedom remain in their ``ground state''. Although the previous reasoning is true in that extreme limit, where frequencies are well apart each other, in actual experimental situations such a limit is hard to reach \cite{Hofferberth}. On the other hand, as we shall argue below, this physical picture is not quite simply obeyed by the Gross-Piatevskii equation. In any, case, we follow here the procedure delineated in Refs. \cite{Trallero, Bao}, in which by properly choosing the sizes of these frequencies, the dynamics along different dimensions can be decoupled. This decoupling allows for an adiabatic approximation such that in the tight direction, namely where $\omega_x = \omega_y \gg \omega_z$, or $\omega_z \gg \omega_x= \omega_y$, along the axis or axes where the frequencies are very large, the condensate wave function can be approximated as that of the ground state harmonic oscillator with the corresponding frequencies. This implies that the non-linear dynamics occurs solely in the directions where the frequencies are small. As briefly discussed below, we consider two ways in which the adiabatic component can be integrated out, one at the level of the energy functional,\cite{Bao} and the other at the level of the GP equation itself.\cite{Trallero} This leads to a small but noticeable difference in the effective non-linear coupling constant in the respectively obtained 1D and 2D dynamical GP equations. In the next subsections we set the equations describing the condensate restricted to 1D and 2D and, then, in the following section we evaluate the validity of the approximations mentioned, by comparing both stationary and dynamical results from the full 3D GP equation with the corresponding versions in the reduced 1D and 2D.

\subsection{Cigar shaped Bose-Einstein condensates}

Let us assume that the frequencies along $x$- and $y$-axes are equal and very large with respect to the frequency along the $z$ direction, that is, $\omega_x=\omega_y = \omega_r  \gg \omega_z$. Thus, following the derivations in  Refs. \cite{Bao, Trallero}, the proposal is that the condensate order parameter $\psi_{3D}(x,y,z,t)$  can be  factorized as,
\begin{equation}
\psi_{3D}(x,y,z,t) \approx \phi_{2D}^{HO}(x,y) \psi_{1D}(z,t) \label{ansatz1D},
\label{Ansatz3D-1D}
\end{equation}
where $\phi_{2D}^{HO}(x,y)$ is the ground state wave function of a 2D harmonic oscillator in the $x-y$ plane with frequency $\omega_r$, and $ \psi_{1D}(z,t)$ is the time dependent wave function describing the dynamics along the $z$-axis. The normalization conditions satisfied by these functions are
\begin{equation}
\int dx \> dy |\phi_{2D}^{HO}(x,y)|^2 =1,
\end{equation}

 \begin{equation}
 \int dz |\psi_{1D}(z,t)|^2 =1.
 \end{equation}
It is evident that the approximation given by Eq.(\ref{Ansatz3D-1D}) replaces the actual cigar-shape condensate by a  ``cylindrical'' one and, as we shall, see this has evident consequences both in their stationary and dynamical properties.\\
 
After the equation for the 2D harmonic oscillator $\psi_{2D}^{HO}(x,y)$ is integrated out (see Refs.~\cite{Bao,Trallero}), the resulting 1D GP equation that describes the time dynamics in the longitudinal $z-$axis is,
\begin{eqnarray}
i \hbar \frac{\partial}{\partial t} \psi_{1D}(z, t)&= &\left [-\frac{\hbar^2}{2m} \frac{\partial ^2}{ \partial z^2} +  \frac{1}{2}m \omega_z^2 z^2 \right.  \nonumber  \\
&&\quad\left. {} + \hbar \omega_r +  \frac{4 \pi \hbar^2 N a}{m}\frac{1}{\pi \alpha l_r^2} \> | \psi_{1D}(z,t)|^2  \right] \psi_{1D}(z,t),
\label{GP1D}
\end{eqnarray}
where $l_r=\sqrt{\hbar/m \omega_r}$  is the natural in-plane harmonic oscillator length along the tight confinement, and the factor $\alpha$ in the effective non-linear coupling is either $\alpha = 2$ if the integration of the transverse modes is performed at the level of the energy functional, \cite{Bao} while $\alpha = 3$ if the integration is done at the 3D GP equation itself. In the following section we will compare the solution of the above approximated equation, as a function of coupling constants $Na$ and the ratio  $\gamma = \omega_z/\omega_r$, with the solution of the full 3D GP. We point out here that  the value of $\gamma$ ranges in the interval $[0.001, 1]$ in typical experimental situations. \cite{Moritz,Hofferberth, Tolra}\\

The stationary solution to the above 1D  GP equation is found by assuming a solution of the type
\begin{equation}
\psi_{1D}(z,t)= e^{-i \mu t/\hbar} \psi_{1D}^s(z) \label{stat1D}
\end{equation}
where $\psi_{1D}^s(z)$ is the stationary solution and $\mu$ is the chemical potential,  
not only of the reduced 1D system, but also of the full 3D gas. We shall return to this point in the comparisons below.

\subsection{Disk shaped Bose-Einstein condensates}

We now consider the equations that describe a time dependent Bose-Einstein condensate lying in the $x-y$ plane. For this purpose we proceed in an analogous way as that described in the previous subsection. We now assume that frequency along the $z$-axis is very large with respect to frequencies in the directions $x$ and $y$, assumed equal, $\omega_x=\omega_y = \omega_r  \ll \omega_z$. Then, in this case, the proposal is to approximate the 3D condensate wave function as
\begin{equation}
\psi_{3D}(x,y,z,t) \approx \psi_{2D}(x,y,t)  \phi_{1D}^{HO}(z),\label{ansatz2D}
\end{equation}
being $\phi_{1D}^{HO}(z)$ the 1D harmonic oscillator for the ground state in direction $z$, having as its natural length $l_z=\sqrt{\hbar/m \omega_z}$. The function $\psi_{2D}(x,y,t)$ describes the time dynamics of a 2D disk shaped condensate situated in the plane $x-y$. The analogous normalization conditions to the cigar shaped condensate become now,
\begin{equation}
\int dx \> dy |\psi_{2D}(x,y,t)|^2 =1 ,
\end{equation}
\begin{equation}
 \int dz |\phi_{1D}^{HO}(z)|^2 =1.
\end{equation}

Again, after integrating out the 1D stationary contribution, one is left with an effective 2D GP equation that describes the dynamics of the condensate:
\begin{eqnarray}
i\hbar \frac{\partial}{\partial t} \psi_{2D}(x, y, t)&=& \left[ -\frac{\hbar^2}{2m} \nabla_{\perp}^2 + \frac{1}{2} \omega_r^2(x^2 + y^2) +\frac{1}{2} \hbar \omega_z  \right.  \nonumber  \\
&&\quad\left. {} +\frac{4 \pi \hbar^2 N a}{m}\frac{1}{\sqrt{\pi \alpha}l_z} |  \psi_{2D} |^2 \right] \psi_{2D}( x, y, t).
\label{GP2D}
\end{eqnarray}
where again, $\alpha  = 2$ or $\alpha = 3$, depending on whether the integration is made at the level of the energy functional \cite{Bao} or the 3D GP equation, following the proposal of Ref \cite{Trallero}. As we discuss below, we will compare the solution of this approximated 2D equation with that of the full 3D GP equation, as a function of $g$ and of the frequencies ratio $\gamma = \omega_z/\omega_r$. Typical values of this ratio are in the interval $[1,10 ^3]$. \cite{Chomaz,Won,Spielman,Petrov} Also, as in the 1D case, the stationary 2D solution is given by,
\begin{equation}
\psi_{2D}(x,y,t)= e^{-i \mu t/\hbar} \psi_{2D}^s(x,y) \label{stat2D}
\end{equation}
with $\mu$ the chemical potential when the gas is in its the ground state, namely, at $T = 0$, both of the reduced 2D system and of the full 3D gas.

\section{Validity tests of dimensional reduction}

The main purpose of this work is to compare the numerically exact solution $\psi_{3D} (\mathrm{\bf r},t)$ to the 3D GP equation (\ref{GP3D}) with the approximated ansatz given by Eqs. (\ref{ansatz1D}) and (\ref{ansatz2D}), respective solutions to the 1D GP equation (\ref{GP1D}) and to the 2D GP equation (\ref{GP2D}), for different values of the ratio $\gamma = \omega_z/\omega_r$ and of the atomic interaction $g$. We base our conclusions on extensive numerical calculations with state-of-the-art parallel computations on GPU processors; see Appendix A for details of the numerical calculations. 

In what follows we use dimensionless units, $\hbar = m = \omega_r = 1$. With this convention, all lengths are adimensionalized with $l_r = (\hbar/m\omega_r)^{1/2}$. We note that the whole problem has 5 parameters, $\hbar$, $m$, $a$, $\omega_z$ and $\omega_r$,  and therefore, with the chosen adimensionalization we are left with two free parameters, namely, the anisotropy ratio $\gamma = \omega_z/\omega_r$ and the dimensionless interaction strength, $g = 4 \pi Na/l_r$. As we now describe, we make comparisons of the following quantities, three of stationary nature and the other a dynamical one. Each quantity is compared for each value of the pair of parameters $(g,\gamma)$ and for the two different  approaches, namely, $\alpha =2$ and $\alpha = 3$.\\

{\bf (I) Chemical potential.} The chemical potential $\mu_{3D}$ of the stationary solution of the 3D GP equation (\ref{GP3D}) is separately compared with the chemical potential $\mu_{1D}$ and $\mu_{2D}$ of the corresponding stationary solutions of the 1D and 2D GP equations (\ref{GP1D}) and (\ref{GP2D}). It is expected that $\mu_{1D} \to \mu_{3D}$ as $\gamma \to 0$, while $\mu_{2D} \to \mu_{3D}$ as $\gamma^{-1} \to 0$.\\

{\bf (II) Stationary transverse wavefunction.} For the 1D case, we compare the exact stationary 3D solution $\psi_{3D}(x,0,0)$, evaluated along the transverse direction $x$, with the product $\phi_{2D}^{HO}(x,0) \psi_{1D}(0)$. This is in order to see if the assumption that  the function  $\psi_{3D}(x,0,0)$ indeed corresponds to that of a harmonic oscillator in the $x$-axis. For the 2D case, the analog is the comparison of the stationary $\psi_{3D}(0,0,z)$ along the longitudinal direction with $\phi_{1D}^{HO}(z) \psi_{2D}(0,0)$.  \\

{\bf (III) Fidelity of the approximated 1D and 2D wavefunctions.} The main assumption behind the 1D and 2D wavefunctions ansatz is an adiabatic approximation. That is, that for sufficiently long times, the approximated wavefunction behave as 1D and 2D solutions when evolved by the full 3D GP equation. To directly test this assumption we take the stationary 1D and 2D solutions, $\phi_{2D}^{HO}(x,y) \psi_{1D}(z)$ and  $\phi_{1D}^{HO}(z) \psi_{2D}(x,y)$ obtained from the respective 1D and 2D GP equations, and evolve them with the {\it full} 3D GP equation. Then, we calculate the ``fidelity'', which is the overlap of the initial state with its evolution as a function of $t$, namely,
\begin{equation}
{\cal C}(t) \equiv \left| \int \psi^*(x,y,z,0) \psi(x,y,z,t) d^3r \right|,
\end{equation}
where $\psi(x,y,z,0) = \phi_{2D}^{HO}(x,y) \psi_{1D}(z)$ for 1D, and  $\psi(x,y,z,0) = \phi_{1D}^{HO}(z) \psi_{2D}(x,y)$ for 2D. If the fidelity ${\cal C}(t) = 1$ for all times, it means that $\psi(x,y,z)$ is a true stationary state. \\

{\bf (IV) Stationary longitudinal wavefunction.} Here we make a very important comparison of the stationary solutions, since this is the direct verification of the stationary solutions of the different GP equations. In the 1D case, we compare the exact stationary 3D wavefunction, along $z$, namely, $\psi_{3D}(0,0,z)$ with $\phi_{2D}^{HO}(0,0) \psi_{1D}(z)$, solution of the 1D GP given by Eq. (\ref{GP1D}). For the 2D case, due to the $x-y$ symmetry, we compare $\psi_{3D}(x,0,0)$ with the solution of the 2D GP equation (\ref{GP2D}), $\phi_{1D}^{HO}(0) \psi_{2D}(x,0)$. \\

Our results are summarized in the panels of Figs. \ref{Figure1} and \ref{Figure2} for the cigar 1D shape, and Figs. \ref{Figure3} and \ref{Figure4} for the disk 2D case.\\

It is interesting to notice that the approximations here considered, \cite{Bao, Trallero} leads us to recognize that the effective coupling interaction of the 3D GP equation $g \sim Na$, is replaced by rescaled effective interaction coupling constants $g^{1D}$ and $g^{2D}$ when the isotropic condensate cloud is squeezed to 1D and 2D respectively, see equations (\ref{GP1D}) and (\ref{GP2D}) where $g^{1D} \sim Na/l_r^2$ and $g^{2D}\sim Na/l_z$, accordingly. It is important to mention here an alternative procedure \cite{Petrov} to find the effective two-body interaction parameter in a 2D harmonic trap, which essentially leads to the case of Ref. \cite{Bao}, namely $\alpha =2$, with an additional logarithmic correction depending on a momentum cutoff. This would lead to an adjustment of the effective coupling only and will not change the main conclusions here presented. 

\section{Comparing the approximated 1D cigar shape BEC with its full 3D solution}

For this case, we performed calculations for nearly 20 values in the range $\gamma = 0.001$ up to $\gamma = 1.0$, and for four cases of the dimensionless interaction strength, $g = 6, 60, 600, 6000$, which, when using values of $^{87}$Rb, correspond to approximately a number of atoms $N = 10^2, 10^3, 10^4, 10^5$. \\

Referring to Figure 1, looking first at the chemical potential, the first column of the panel, we plot the three obtained values $\mu_{3D}$, in dimensionless units,
the chemical potential of the full 3D solution, and the values $\mu_\alpha$, $\alpha = 2$ and 3, corresponding to the two versions of the 1D solutions. As expected, in all cases, it is clear that as $\gamma \to 0$, the three of them reach the same value. 
We observe that for intermediate and small values of the interaction the value of the chemical potential is in good agreement with its 3D version for not very small values of $\gamma$, the case $\alpha = 3$ fairing better than $\alpha=2$. In particular, for $g = 6$ ($N = 10^2$), the chemical potential $\mu$ is essentially independent of the anisotropy $\gamma$. The conclusion is that the non-linear term is just a perturbation to an uncoupled 3D harmonic oscillator. In other words, dimensionality plays a role only as the interaction is increased. In this regard, the interesting and notorious observations is that as $\gamma \to 1$, namely, reaching the isotropic case, the expected discrepancy becomes clear as the strength $g$ is increased. This is consistent with the other tests, as we now discuss.\\
\begin{figure}[h]
\begin{center}
\includegraphics[width=16cm, height=14cm]{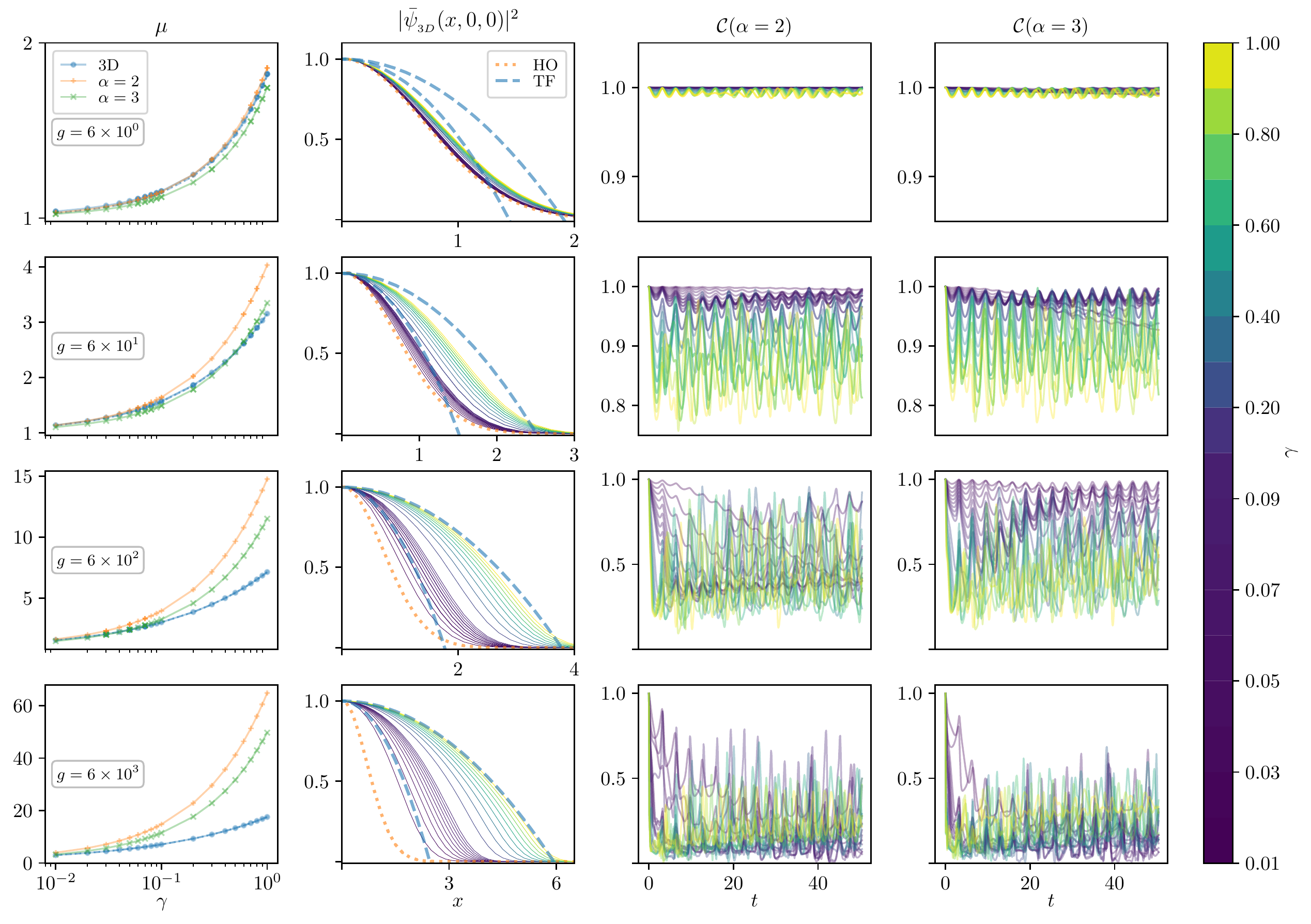}
\end{center}
\caption{Stationary and dynamical properties for the cigar 1D shape condensate. From left to right, panels correspond to the chemical potential $\mu$, the density profile along the transverse $x$ direction $\vert \overline{\psi}_{3D}(x,0,0)\vert ^{2}$ (solid lines) normalized to its value at the origin, and the time dependent fidelity ${\cal C}(t)$, associated to approximations $\alpha = 2$ and $\alpha = 3$. From top to bottom different panels correspond to dimensionless values of the interaction $g=6,60,600,6000$ which for $^{87}$Rb correspond to $N = 10^2$,  $10^3$, $10^4$ and $10^5$, respectively. In the second column of figures the dotted orange lines correspond to the 2D harmonic oscillator ground state wave function, while the blue dashed lines refer to the Thomas-Fermi approximations for the lowest and highest values of $\gamma$. The color code for the values of $\gamma$ is indicated in the right column.}
\label{Figure1}
\end{figure}

The next comparison is the transverse part of the normalized condensate density profile, $\vert \overline{\psi}_{3D}(x,0,0)\vert ^{2}$ (solid lines), second column in the panel of Figure \ref{Figure1}. We point out that the ansatz to reduce the dimensionality is the assumption that in the $(x,y)$ coordinates, the wavefunction is that of the ground state of a 2D harmonic oscillator, see Eq. (\ref{ansatz1D}). In all figures there is a dotted line (orange), labeled (HO), and two dashed lines (blue), labelled (TF). The former is the plot of the harmonic oscillator density in the  $x-$direction of the ansatz, normalized to its value at $x=0$, i.e. $\vert {\phi}_{2D}^{HO}(x,0) \psi_{1D}(0)\vert ^{2}$, while the latter are the Thomas-Fermi approximation for the extreme values $\gamma = 0.01$ and $\gamma = 1.0$. Here we recall that the TF approximation is the limit when the kinetic energy term (proportional to the laplacian) in the GP equations, compared with the non-linear term, can be neglected; this yields, for a generic harmonic potential,
\begin{equation}
\rho_{TF}(r) = \frac{1}{g} \left(\mu - \frac{1}{2} m \omega^2 r^2 \right)
\end{equation}
if the right-hand-side is positive, and $\rho_{TF}(r) = 0$ if negative. We note that for $g = 6$, the HO ansatz is ``too'' good since, in agreement with the results of the chemical potential, the transverse wavefunction is indeed a HO one, but for all values of $\gamma$, just corroborating that the interaction term is a perturbation.  As $g$ is increased, and for $\gamma \to 1$, the transverse part tends to the TF solution. However, for small $\gamma$ the approach to the HO profile is too slow, deteriorating even more for larger values of $g$.  The overall conclusion is that as the interaction is large enough, $g \ge 60$, the 2D harmonic oscillator wave function is a reasonable good approximation for $\gamma$ sufficiently small, namely, for $g = 60$, $\gamma \le 0.1$ and for $g = 600$, $\gamma \le 0.001$, and so on.  At this stage it is worth mentioning that there are two other variational approaches that lead to approximated 1D and 2D GP-like equations \cite{Reatto,Mateo} which show two different regimes, a weakly and a strong interacting ones.  The former coincides with the 1D and 2D approaches analyzed here but the latter seemingly describes a different regime where interactions are more relevant. In such a regime, those approaches lead to {\it transverse} profiles close to a Thomas-Fermi one, rather than to a gaussian shape. What we find, however, with our exact 3D calculations, is that the strong interaction regime may apply when the anisotropy is very small only, namely, when $\gamma \sim 1$, since it is in this limit when the TF approximation for the transverse part is approached. Although those approximations \cite{Reatto,Mateo} may be useful in some particular cases, it is our conclusion that for condensates nearly isotropic one should remain within a numerical solution of the full 3D GP equation.\\

The third and fourth columns in Fig. \ref{Figure1} show the fidelity of the stationary states, for $\alpha = 2$ and $\alpha = 3$, when evolved with the full time-dependent 3D GP equation, Eq.(\ref{GP3D}). These results show, again, that for small $g$ the fidelity is better than for large $g$, corroborating that the non-linear term amounts to a perturbation. For $g = 600$ and $g = 6000$, the results are not very encouraging, signaling that the adiabatic ansatz is not satisfactory for all the considered values of $\gamma$, it appears that values of $\gamma$ smaller than $0.001$ should be considered. At first sight, within this test, it seems that the 1D limit is better achieved for $g = 60$ and $\gamma \approx 0.01$. But we still need to cross this conclusion with the actual longitudinal behavior at 1D, shown in Figure 2 below. It is of interest to mention that the wavefunction $\psi_{1D}(z) $ in the case $\alpha = 3$ \cite{Trallero} appears a bit better than $\alpha = 2$ \cite{Bao}, but with no major differences. We also point out that the oscillations shown by all the time evolutions of the fidelity, correspond to the frequency of the trap, indicating a kind of ``revival'' of a better agreement every period of the oscillations at $\omega_r$. These oscillations are a breathing mode of the transverse modes induced by the initial approximation that the shape of the condensate is actually a cylinder than a cigar-like one.\\
\begin{figure}[h]
\begin{center}
\includegraphics[width=16cm, height=14cm]{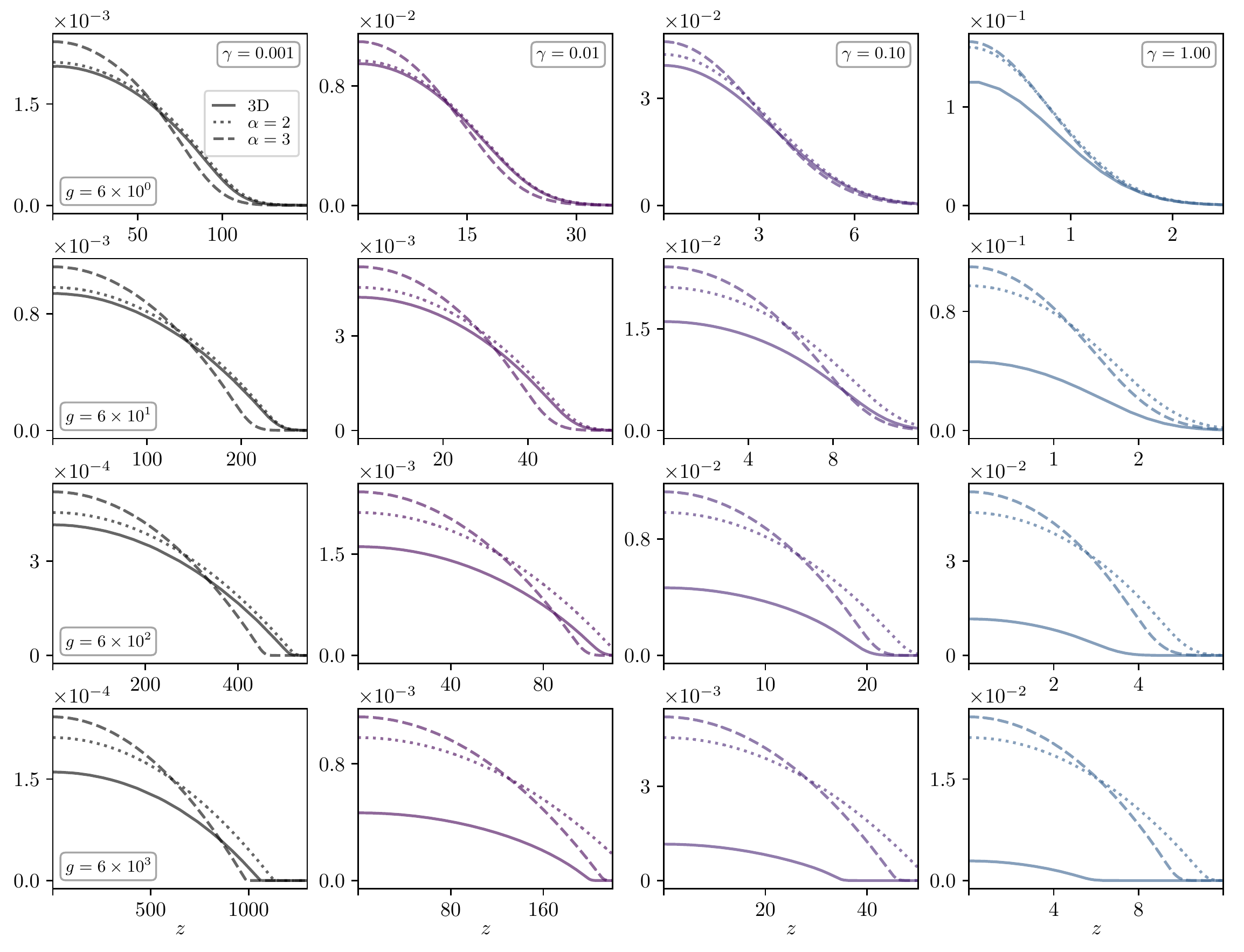}
\end{center}
\caption{Normalized density profiles $|\psi(0,0,z)|^2$ of condensate wavefunctions along the longitudinal direction $z$, associated to different values of the ratio $\gamma$ and dimensionless coupling interaction $g$, from top to bottom $g=6,60,600,6000$, and from left to right $\gamma = 0.001, 0.01,0.1, 1$ respectively. Continuous, dotted and dashed lines correspond to 3D GP solution $\psi_{3D}(0,0,z)$ and approximations $\psi_{1D}(z)$ of Refs. \cite{Bao, Trallero}, respectively.}
\label{Figure2}
\end{figure}

Figure \ref{Figure2} shows comparisons of the 3D condensate density profiles along the longitudinal $z-$direction, $\psi_{3D}(0,0,z)$ with the corresponding 1D solutions $\phi_{2D}^{HO}(0,0) \psi_{1D}(z)$, for $\alpha = 2$ and $\alpha =3$, for the same cases of $g$ as in Fig. \ref{Figure1}, and for the selected values $\gamma = 0.001, 0.01, 0.1, 1$. Note that in Fig. \ref{Figure1} the lowest value of $\gamma$ is 0.01. Once more, the best agreement for the order parameter is for the smallest $g = 6$, yet, the 1D profiles are very close to those of a harmonic oscillator. The expectation is that, in order to see GP behavior, the profiles should be close to a TF type of profile. This is certainly corroborated as $g$ is increased, namely, that while the agreement between the 1D approximations and the full 3D calculations may not be very good, all of them tend to a TF profile. Crossing now the results of this figure with those of Fig. \ref{Figure1}, one finds again that for $g = 60$ and $\gamma = 0.01$ there is a good agreement with TF behavior, specially for
$\alpha = 2$. The purpose of including the very small value $\gamma = 0.001$ here, allows us to conclude that a similar agreement is found for $g = 600$ and $\gamma = 0.001$, and for $\alpha = 2$. This trend suggests that for $g = 6000$, a very large value, the agreement should be found for values near $\gamma = 0.0001$. This latter value is beyond any experimental realizations reported. Moreover, as we indicate in Appendix A, such a small value of $\gamma$ cannot be calculated numerically here with the same accuracy used for the other calculations.
It is important to emphasize that a good approximation of the chemical potential can be reached for a certain range of the ratio $\gamma$ even for large values the interatomic parameter $g$ (see the first column in Fig. \ref{Figure1}).\\

An overall conclusion here, to be further discussed in the final section, is that, although the non-linear 1D behavior is indeed reached for smaller values of $\gamma$ as $g$ is increased, there is a crossover value below which, dimensionality plays no role. 

\section{Comparing the approximated 2D disk shape BEC with its full 3D solution}

In this section we now compare the GP solution associated to a disk shaped condensate lying in the $x-y$ plane, to the solution in 3D space. As in the previous 1D-3D comparison, we consider four cases of the interaction strength, $g = 6, 60, 600, 6000$,  which for $^{87}$Rb, correspond to a number of atoms $N = 10^2, 10^3, 10^4, 10^5$. We also compare the two approximated versions of the 2D GP equations, Eq. (\ref{GP2D}), with $\alpha = 2$ and $\alpha = 3$. Since we keep the same notation for the frequency ratio $\gamma = \omega_r/\omega_z$, now $\gamma \ge 1$, hence, we present our results as a function of $\gamma^{-1}$, ranging from 0.01 to 1. We proceed similarly as before discussing stationary and dynamical properties.\\
\begin{figure}[h]
\begin{center}
\includegraphics[width=16cm, height=13cm]{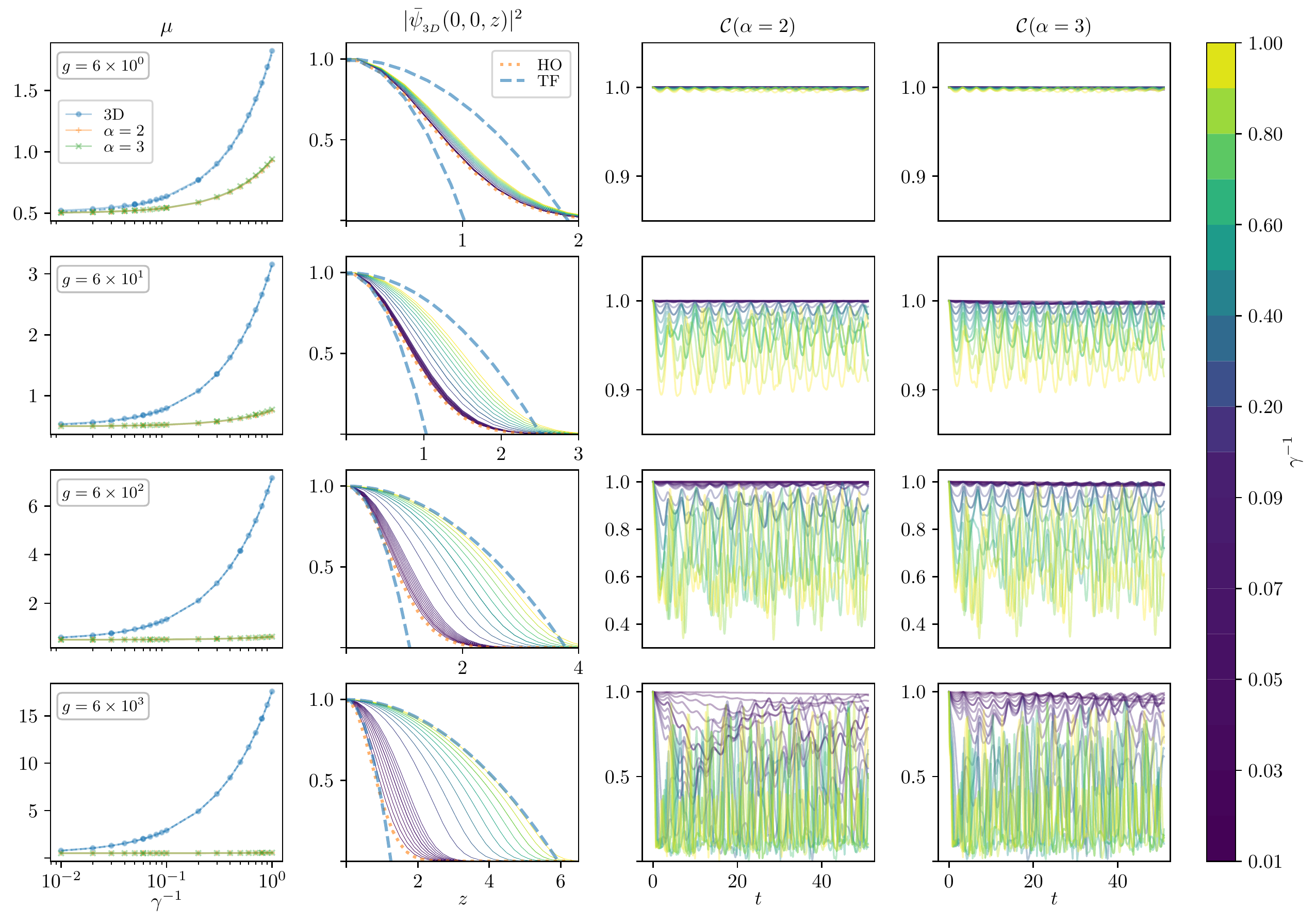}
\end{center}
\caption{Stationary and dynamical properties for the disk 2D shape condensate. From left to right, panels correspond to the chemical potential $\mu$, the density profile along the transverse $z$ direction $\vert \overline{\psi}_{3D}(0,0,z)\vert ^{2}$ (solid lines), normalized to its value at the origin, and the time dependent fidelity ${\cal C}(t)$, associated to approximations $\alpha = 2$ and $\alpha = 3$. From top to bottom, different panels correspond to values  (in reduced units) $g=6,60,600,6000$ which for $^{87}$Rb correspond to $N = 10^2$,  $10^3$, $10^4$ and $10^5$, respectively. In the second column of figures the dotted orange lines correspond to the 1D harmonic oscillator ground state wave function, while the blue dashed lines refer to the Thomas-Fermi approximations for the lowest and highest values of $\gamma$. The color code for the values of $\gamma^{-1}$ is indicated in the right column.}
\label{Figure3}
\end{figure}

In Fig. \ref{Figure3}, first column, the chemical potential is plotted as a function of $\gamma^{-1}$ and, as expected, as $\gamma^{-1} \to 0$ the 3D and the two 2D cases converge to the same value. We note that the values of the chemical potential for the 2D versions are essentially the same, in contrast with the 1D situation. Although the 2D approximation should only be used as $\gamma^{-1} \to 0$, one would like to see how small should actually be this parameter. We observe that, even for small $g$, the discrepancy between 3D and 2D is always clear for $\gamma^{-1} \to 1$, becoming even more evident as $g$ is increased, reaching a situation where the chemical potential remains almost constant for all values of $\gamma^{-1}$. Hence, according to the chemical potential, 2D approximations can be considered appropriate until $\gamma^{-1} \approx 0.01$. This is too stringent, with respect to the following tests.\\
\begin{figure}[h]
\begin{center}
\includegraphics[width=16cm, height=15cm]{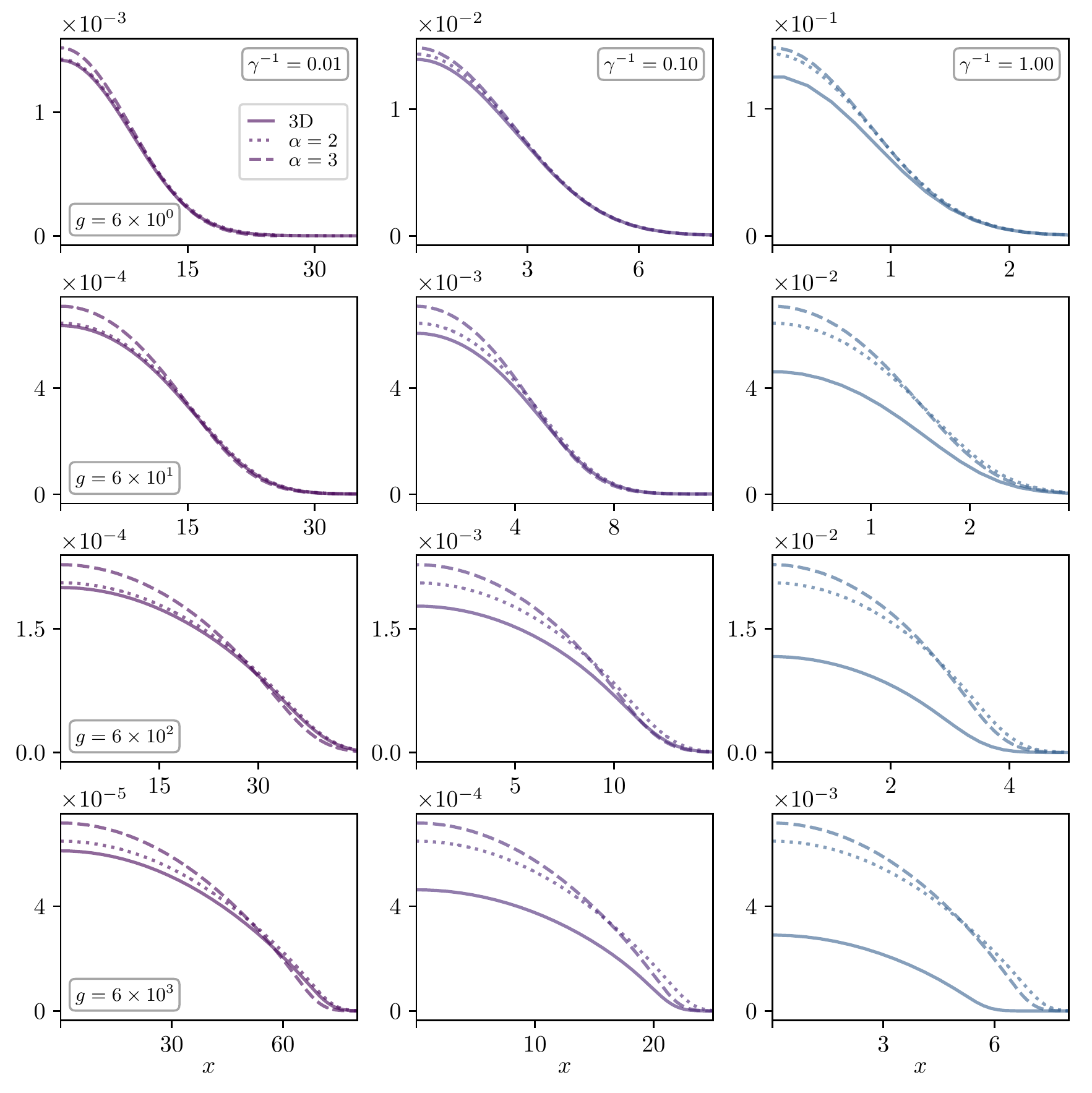}
\end{center}
\caption{Normalized density profiles $|\psi(x,0,0)|^2$ of dimensionless condensate wavefunctions along the direction $x$, associated to different values of the ratio $\gamma$ and coupling interaction $g$, from top to bottom $g=6,60,600,6000$, and from left to right $\gamma^{-1} = 0.01,0.1, 1$ respectively . Continuous, dotted and dashed lines correspond to 3D GP solution $\psi_{3D}(x,0,0)$ and approximations $\psi_{2D}(x,0)$ for $\alpha=2$ and $\alpha = 3$,  respectively.}
\label{Figure4}
\end{figure}

The second column of panels in Fig. \ref{Figure3} shows the transverse dependence of the density profiles, which in this case is along the $z$ coordinate, the disk being on the $x-y$ plane. The ansatz Eq.(\ref{ansatz2D}) in both approximations is, again, the ground state wavefunction of a 1D harmonic oscillator of frequency $\omega_z$. This is shown with a (orange) dotted line. The curves plotted correspond to the different 3D calculations of the wavefunction along the $z$ direction, but all normalized by the wavefunction value at $(0,0,0)$ to better appreciate the differences with respect to the gaussian HO and the TF profiles. The dashed (blue) lines are the Thomas-Fermi approximation for the $\gamma^{-1} = 1.0$ and $\gamma^{-1} = 0.01$ profiles. Similarly to the 1D case, the dependence on $g$ is clear: for small $g \sim 6$, most of the transverse profile approach essentially the gaussian profile, almost independently of the value of $\gamma^{-1}$. That is, TF behavior is never seen, indicating that this system is close to a 3D uncoupled HO. On the other hand, as $g$ is increased, the profiles are close to the gaussian one for small $\gamma^{-1} $, becoming TF as the system becomes isotropic. A contrast to the 1D case, an reiterated by the tests below, is that for $\gamma^{-1} \le 0.1$ the transverse dependence along $z$ tends very nicely to the gaussian profile for essentially all values of $g$, yielding a good validity to the proposed ansatz.\\

Another very interesting aspect of the 2D approximations is found in the time evolution of the fidelity, shown in the third and fourth columns of Fig. \ref{Figure3}, where we observe that for small $\gamma^{-1} \le 0.1$, the 2D calculations, specially for $\alpha = 3$ are quite stationary for all values of $g$. We recall that in the 1D case this type of agreement deteriorates as $g$ increased. Putting together the two previous tests, chemical potential and wavefunction transverse dependence, with the fidelity time evolution, one can conclude that $\gamma^{-1} \sim 0.01-0.1$ are enough to reach a 2D disk condensate, although the 2D GP physics is better seen as $g$ increases, just as in the 1D case.\\

The last test, shown in the panels of Fig. \ref{Figure4}, corroborate the conclusion of the last paragraph, namely, that $\gamma^{-1} \le  0.1$, and $g \ge 60$, describe an interacting condensate in 2D. This can clearly be seen in the first and second columns corresponding to $\gamma^{-1} = 0.01$ and $0.1$, where the 2D profiles are very close to the 3D one, specially the case $\alpha = 2$ in which the dimensional reduction is performed at the level of the energy functional.\cite{Bao}\\

\section{Final Remarks}

As expressed in the Introduction, physical properties of many-body systems tend to have a strong and rich dependence on dimensionality. This motivates the design of experiments that reduce dimensionality, from our 3D world, to test and understand those differences and to explore different physical phenomena. Bose-Einstein condensation is not an exception and, while the Gross-Pitaevskii approximation does not encompass all its physical behavior since it describes the superfluid at zero temperature only, it serves nevertheless as a tool to test the validity of the, many times simply assumed, reduced dimensionality of the GP equation. This is the main purpose of the present study.\\

In Section II we briefly presented the GP equations that can be obtained from an adiabatic approximation, in the limit $\gamma \to 0$ in 1D and $\gamma^{-1} \to 0$ in 2D. \cite{Bao,Trallero} Those procedures differ slightly in their final obtention of the corresponding GP equations, and the only discrepancy is that the effective 1D and 2D atomic interaction parameter $g$ are the factors $1/\pi \alpha$ in 1D and $1/\sqrt{\pi \alpha}$ in 2D, see Eqs. (\ref{GP1D}) and (\ref{GP2D}). There is, however, an additional subtle point in the dimensional reduction that we address now. The physical picture of such a reduction is that from an interacting Bose gas in a 3D harmonic potential, one or two of the dimensions are ``squeezed" by largely increasing the corresponding trap frequencies. As a consequence, the energy levels along the squeezed dimensions become so separated that the system remains in its lowest energy state along those directions, and the non-linear interacting dynamics occurs in the remaining 1D or 2D degrees of freedom. The point we make here is that the previous picture is not quite represented by the reduction in the GP equation. That is, because the equation mathematically has five parameters, $\hbar$, $m$, $g$, $\omega_r$ and $\omega_z$, three are used for dimensions, say $\hbar$, $m$ and $\omega_r$. Thus, the change in dimensions is obtained by the dimensionless parameter $\gamma = \omega_z/\omega_z$, such that $\gamma \ll1$ can be obtained by keeping the value of $\omega_z$ and increasing $\omega_r$, namely squeezing the transverse dimension, or by enlarging $\omega_z$. But in the latter case the ``energy levels'' in the transverse directions were not affected, yet, as $\gamma \to 0$ the dimension reduction is obtained. As a matter of fact, in our numerical code we kept $\omega_r$ and reduce $\omega_z$ to keep the same level of accuracy in all the calculations. The obvious conclusion is that GP is insensitive to either change in the frequencies and therefore, the physics of the reduction is a different one from the naive picture. At the level of the GP equation, the answer of the dimension reduction lies in the non-linearity of the equation, which physically represents the interatomic interactions: the anisotropy of the external potential, $\gamma \to 0$ or $\gamma \to \infty$, results in a decoupling of the interaction in the {\it relatively} smaller potential lengths, independently of their actual value. However, as clearly indicated by the calculation, the true reduction from a 3D to a 1D or 2D system is not relative, it occurs for values of the interaction larger than $g \sim 10$, the transition being a crossover rather than a sharp one. For smaller values of $g$, the system is essentially a harmonic oscillator decoupled in its three dimensions, the non-linear interaction being a perturbation.\\

Comparing and summarizing Figs. \ref{Figure1}-\ref{Figure4}, the reduction for 1D is more delicate than in 2D. It may be that reducing two dimensions is harder than reducing only one. For the 1D case, the best achievement is for $g \approx 60$ and $\gamma \approx 0.01$, and the expectation is that an increase in a order of magnitude for $g$ requires a decrement in $\gamma$ for also an order of magnitude approximately, namely $g = 600$ requires $\gamma \approx 0.001$. Nevertheless, considering purely global properties, such as the chemical potential, the approximation do much better in 1D than in 2D for values of the anisotropy parameter $\gamma$ not so small; this is important to keep in mind as being a typical consequence of variational approaches. Regarding experiments, it seems that the strong limit $\gamma \sim 0.001$ for $g \sim 600$ are not easily achieved, however, for the 2D reduction, the bounds are better. That is, for $\gamma^{-1} \sim 0.01 - 0.1$ and for $g \ge 60$ all cases are a good 2D system. In particular, a promising candidate is $g = 6000$ ($N = 10^5$) and $\gamma^{-1} = 0.01$, for this is typical of actual experimental scenarios. Regarding the two types of reductions, it is not definite which one is more accurate, since $\alpha = 3$ is better in the chemical potential and the fidelity tests, while $\alpha = 2$ appears more reliable in the prediction of the stationary density profile in the reduced dimensions. 

\section*{Appendix: Details of the GP numerical solutions}

To perform the tests described above, we numerically solve the stationary and time-dependent GP equations in their 1D, 2D and 3D versions. We use dimensionless variables $\hbar = m = \omega_r = 1$, yielding two dimensionless free parameters $\gamma = \omega_z/\omega_r$ and the interaction strength $g$. For purposes of potential comparisons with current experiments, we use the value of the scattering length of $^{87}$Rb, $a \approx 50 \times 10^{-8}$ cm.  This yields the dimensionless values that we use, $g = 6, 60, 600, 6000$, which correspond to number of particles, $N = 10^2, 10^3, 10^4, 10^5$.\\

The stationary GP equations are solved with the method of imaginary time evolution \cite{Zamora1,Zeng,Bao3} and the time-dependent GP one with a fourth-order Runge-Kutta (RK-4) evolution. \cite{Taha,Balac,Caplan} The laplacians in the equations can be calculated with finite differences or with spectral methods yielding the same results. The codes are performed in state-of-the-art parallel GPU processors. We use a system with four GPU Nvidia GTX 1080  with  a total amount of 10,240 CUDA cores achieving a considerable saving of computing time. To perform GPU programming of the GP equations we use Python libraries such as PyCUDA, Scikit-Cuda, pyFFT and Scipy. \cite{pycuda,skcuda,pyfft,scipy} \\

The most demanding calculations are those in 3D and in order to keep the same level of accuracy for the different values of $\gamma$, we use cartesian grids in all dimensions, with a constant value $\Delta l$ for every length interval; this interval typically take the values 0.05, 0.1 or 0.2. Our 3D grids range from $512 \times 512 \times 512$ for isotropic clouds up to very elongated ones $64\times 64\times 8192$ for 1D cigar-shaped condensates and $1024 \times 1024 \times 64$ for very flat disk-shaped ones. The time steps imaginary for the time evolution are $d\tau=0.002,$ and $0.001$, and for real time evolution we consider $\Delta t=0.001$ and $0.0001$. As expected, we recover analytical solutions for the HO when $g=0$. To test the correctness of all stationary states we perform an RK-4 time propagation of each studied case for $t=50$ units of time, and verify that the energy and normalization remain constant within a relative error around $10^{-5}$. Furthermore, as a double check, we analyze distinct simulation boxes for some cases, obtaining the same wave functions and chemical potentials.
The same RK-4 evolution is used for the calculation of the fidelity overlaps. We point out that the total time evolution of 50 time steps plus relax iterations to find stationary states, were equivalent of nearly 100 hours of GPU processor. Thus, considering the full set of values of $\gamma$, as well as different numbers of $N$ analyzed, the total time in the numerical calculations here studied demanded nearly 3600 hours of a GPU processor.

\ack
This work was partially funded by grants IN105217 DGAPA (UNAM), 255573 (CONACYT) and LN-232652 (CONACYT). G.A.D-C acknowledges a scholarship from CONACYT. C.T-G acknowledges support from the Brazilian Agency CNPq.

\end{document}